\documentclass[12pt]{amsart}

\title[Nonholonomic systems with conformally symplectic reduction]{Non-holonomic systems with symmetry \\ allowing a conformally symplectic reduction} 

\author[Rios \& Koiller]{Pedro de M. Rios \& Jair Koiller}  
\address{Laborat\'orio Nacional de Computa\c{c}\~ao Cient\'{\i}fica, Av. Get\'ulio Vargas 333, Petr\'opolis, 
RJ 25651-070, Brazil. \ {\it Present address (Rios): Dept. of Mathematics, Univ. of California, Berkeley.}}  
\email{prios@math.berkeley.edu , jair@lncc.br}

 1

\def\R{{I\!\!R}}

\def\be{\begin{equation}}
\def\ee{\end{equation}}
\def\bq{\begin{eqnarray}}
\def\eq{\end{eqnarray}}
\def\beq{\begin{eqnarray*}}
\def\eeq{\end{eqnarray*}}

\def\R{\hbox{\bf R}}

\begin{document}

\thispagestyle{empty}

\vspace{1cm}

\begin{abstract}
Non-holonomic mechanical systems can be described by a degenerate 
almost-Poisson structure \cite{SM} (dropping  the Jacobi identity) 
in the constrained space. 
If enough symmetries transversal to the constraints are present,
the system reduces to a nondegenerate almost-Poisson structure on a 
``compressed'' space. Here we show, in the simplest non-holonomic systems, that
in favorable circumnstances the compressed system is conformally symplectic, 
although the ``non-compressed'' constrained  system never admits a Jacobi structure
(in the sense of Marle et al. \cite{Marle}\cite{Marle2}). 
\end{abstract}


\maketitle 

\section{Introduction}

We adopt in this work the  view of writing the equations of a non-holonomic systems in terms of an almost-Poisson bracket on a constrained manifold $P \subset T^*Q$,
introduced by van der Shaft and Mashke \cite{SM}. See also  Cantrijn et al. \cite{Cantrijn}
for more recent developments. 

In this note we add a new twist to the simplest example of a  non-holonomic system, the contact system in $Q = \R^3$ (see
eg. \cite{Bates}), namely:

{\it After
performing the reduction by the transversal $\R^1-$symmetry, we observe that, 
for some metrics,   
the  reduced (``compressed'') 
almost-Poisson bivector admits  a conformal 
symplectic structure, ie, a special Jacobi structure \cite{Marle}\cite{Marle2}. In contrast, the ``non-compressed''  constrained system never admits a Jacobi structure.}  

The  examples  point to the fact  that, in quite favorable circunstances, the reduced system can be studied by symplectic techniques   
(for instance, when internal symmetries are present, integrability can be achieved by Marsden-Weinstein procedure, which holds in the
conformally symplectic setting, see \cite{Haller}). The examples also show that, generally, non-holonomic systems are non-Jacobi systems 
(in the sense of Marle et al.  \cite{Marle}\cite{Marle2}).  

Hopefully our observations can help attracting more interest to 
investigations on the 
geometrical properties of almost-Poisson bivectors  which naturally describe non-holonomic dynamics on the constrained  
 submanifold  
of the original cotangent bundle.

\section{The contact non-holonomic system}

Consider a non-holonomic systems in $Q = \R^3$ having the 
constraint 
\begin{equation} \label{contact}
\dot{z} - x \dot{y} = 0  \,\,.
\end{equation}
The admissible sub-bundle $E$ is the union of the horizontal spaces for the connection $1$-form    
$\omega = dz - xdy$  on the (trivial) bundle $G = \R^1 \hookrightarrow Q = \R^3 \rightarrow S = \R^2$,
with curvature $d\omega = - dx\wedge dy$, where the $\R^1$ action is 
the usual translation on the $z$-fibers.  

For  motivation, consider the javelin,   a rod of mass $m $ and 
moment of inertia $I  $ moving on a vertical
plane $(y,z)$ in such a way that it always remains tangent to its 
trajectory.  If $\varphi$
is the angle with the horizontal, then $ dz = \tan(\varphi) \,dy $, and 
if we introduce
the change of variables $ x = \tan \varphi$, then 
$ T = \frac{1}{2} \, [\, m (\dot{y}^2 + \dot{z}^2) +  I 
\dot{\varphi}^2 ] = 
 \frac{1}{2} \, [\, m (\dot{y}^2 + \dot{z}^2) +  I \frac{\dot{x}^2}{(1 
+ x^2)^2} ]$ . 
Choose units so that $m=I=1$.  Under the assumption of small angles, $ 
\varphi \approx x$, we get  
\begin{equation} \label{everybody}
  L = T = \frac{1}{2}\,( \dot{x}^2 + \dot{y}^2 + \dot{z}^2 ) \ .
\end{equation}
According to our previous work (\cite{Koiller}) system (\ref{contact},\ref {everybody}) is ``z-Caplygin'' and thus can be reduced to $TS = \{(x,y,\dot{x},\dot{y})\}$ with lagrangian 
\begin{equation} \label{reducedlagrangian}
  \overline{L} = \frac{1}{2}\,( \dot{x}^2 + (1 + x^2)\, \dot{y}^2) \ .
\end{equation}
with an external gyroscopic force added to it. 
Actually it can be also interesting to regard it as
a y-Caplygin system $ dy = \cot(\varphi)\, dz = (1/x) dz$ , since we may 
want to add the gravitational
potential $ V =  g z $.

\section{Almost-Poisson brackets via moving frames} 

The giroscopic force can be concealed in an almost Poisson
bracket in the constrained manifold  $P \subset T^*Q$ via the dynamical equation 
$$   \dot{x} = \{ x, H \}_P  $$
where $H$ is the hamiltonian, and $P \subset T^*Q$ is the Legendre transform of the
constraint subbundle $E \subset TQ$, as defined below.

Let $e_J = e_{LJ} \partial/\partial q_L$ be a moving frame on  $Q$ such that
the first $m$ vectors (labelled by lowercase latin indices) generate $E$.
By a direct calculation (\cite{SM}, equation(19)) van der Shaft
and Maschke verified that the brackets are given by
$$
\{q_I, q_J \} = 0\,,\,
\{q_I, \tilde{p}_J \} = e_{LJ}(q) = dq_I \cdot e_J\, \,\,\,\,(e_J = 
e_{LJ} \partial/\partial q_L)
$$
\begin{equation} \label{SM}
 \{\tilde{p}_I,\tilde{p}_J\} = - p_q \cdot [e_I,e_J] \equiv R_{IJ}
\end{equation} 
where   $p_q$ is evaluated on $P$. 
For a geometric interpretation  and  simple derivation of
these formulas using the moving frame method see our paper \cite{KR}. 
Here's an outline: 

Choose an {\it adapted}  moving 
frame  to the constrained distribution $E\subset TQ$  that is, 
consider a complete set of vector fields 
$e_i$ , $e_{\alpha}$ where $e_i(q) \in E_q$. The greek labels $\alpha$ denote  $e_{\alpha} \notin E_q$. Denote the dual  1-forms by  $\epsilon_i$ , $\epsilon_{\alpha}$. 
We shall denote by  uppercase latin indices $e_I$ , $\epsilon_I$ the complete dual set
in $TQ$ and $T^*Q$.  The 
canonical $1$-form on $T^*Q$ writes as 
\begin{equation}\label{pdq}  
p \,dq = u_I\epsilon_I 
\end{equation} 
where $u_I$ define new coordinates on each $T^*_qQ$. Now, 
$$ d(pdq) = du_I\wedge\epsilon_I + u_Id\epsilon_I  \ . $$
The last term vanishes on vertical vectors. Moreover 
$$  u_Id\epsilon_I(e_J,e_K) = u_Ie_J(\epsilon_I(e_K)) - 
u_Ie_K(\epsilon_I(e_J)) - u_I\epsilon_I[e_J,e_K] = - u_I\epsilon_I[e_J,e_K] $$
which gives us the symplectic matrix on $T^*Q$  as 
\begin{equation} \label{Omega}
[\hat{\Omega}]_{{\rm moving}\,{\rm frame}} = \left(
\begin{array}{ll}   R & -I_n\\   I_n & \,\,\,\,0_n
\end{array} \right)
\end{equation}
and its inverse, the Poisson matrix on $T^*Q$  as $[\hat{\Lambda}]_{{\rm moving}\,{\rm coframe}} = [\hat{\Omega}]^{-1}_{{\rm moving}\,{\rm coframe}} \ $ , where 
\begin{equation} \label{R} 
R_{JK} = - u_I
\epsilon_I[e_J,e_K] = -p_q \cdot [e_J,e_K] \  . 
\end{equation} 
The moving coframe for $[\hat{\Lambda}]_{{\rm moving}\,{\rm coframe}} \ $ is $\epsilon_I, du_I$.  The moving
frame for (\ref{Omega}) is its dual. Caveat: this basis contains the vertical
vectors $\partial/\partial u_I$ and lifted $e^*_I = e_I + vertical$ (see \cite{KR}). \ 
From this we obtain the   matrix of almost-Poisson brackets in $P\subset T^*Q$ by 
directly substituting the $u_{\alpha}$ by the $u_i$ (and q's) via the 
defining equation for $P$ : 
\begin{equation} \label{defP}
\partial H / \partial u_{\alpha} = 0 
\end{equation}
which is a consistency requirement for any constrained (holonomic or 
non-holonomic) dynamical system on $Q$ , see \cite{KR}.

\section{The contact almost-Poisson structure}

In the case of the contact system, let us begin by 
taking as moving co-frame the set 
\begin{equation} \label{coframe}
\epsilon_{1} = dx\,, \, \epsilon_2 = dy\,,\, \epsilon_3 = \omega = dz - 
x dy
\end{equation}
that is dual to
\begin{equation} \label{frame}
e_1 = \partial/\partial x \,, \, e_2 = \partial/\partial y +
 x \partial/\partial z \,,\, e_3 = \partial/\partial z \,\,.
\end{equation}
Note that    
\begin{equation} \label{Heisen}
[e_1, e_2] = e_3 \ , \ [e_1, e_3] = [e_2, e_3] = 0 \ ,  
\end{equation} 
 so that we have the 3-dimensional Heisenberg algebra. The identity (\ref{pdq}) yields   
\begin{equation} \label{change} 
u_1 = p_x   \,,\, u_2 =  p_y + x p_z     \,,\,  u_3 = p_z \ , 
\end{equation} 
hence the Hamiltonian is 
\begin{equation} \label{hamiltonian} 
H = \frac{1}{2}( p_x^2 + p_y^2 + p_z^2) = 
\frac{1}{2}( u_1^2 + (u_2 - xu_3)^2 + u_3^2) \ .
\end{equation}  

Now, the Legendre transform  of the admissible sub-bundle $\mathcal{E} 
\subset TQ $ is
given by  (\ref{defP}) as 
$  \partial H / \partial u_3 = 0  $  
so that the constrained manifold $P$ is 
\begin{equation} 
P =  \{ (x,y,z, u_1, u_2, u_3)\,|\, u_3 = \frac{x u_2}{ 1 + x^2}\,\,  
\} = \{ (x,y,z, p_x, p_y, p_z)\,|\, p_z =  x p_y\,\,  \} \ . 
\end{equation}
As moving basis for $ T^*P$ we take $dx, dy, dz - x dy, du_1, du_2$ and 
the corresponding $ 5 \times 5 $ matrix of Poisson
Brackets is given by
\begin{equation} \label{QP}   [\Lambda]_{\rm moving}  =   \left(
\begin{array}{lll}   0_{2\times 2} & 0_{2 \times 1}
& I_{2\times 2} \\ 0_{1 \times2} & 0_{1 \times 1 } & 0_{1 \times 2} \\   
-I_{2 \times 2} &  0_{2 \times 1}  & R_{2 \times 2} \end{array} \right)
\end{equation}
where $R_{2\times 2}$ is the antisymmetric matrix with
\begin{equation}  R_{12} = -  p_q [e_1,e_2] = - u_3 = - \,\frac{x u_2}{ 1 + x^2} \ . 
\end{equation}
Here, the vanishing of the middle row and column means of course that 
\begin{equation} \{ dz - x dy, H \} = 0  \,\,\, \Longrightarrow \,\,\, \dot{z} - x 
\dot{y} = 0 
\end{equation}
which gives the differential equation for $z$.

\section{The compressed system}

Deleting the middle column  and row of (\ref{QP}), we obtain a 
non-degenerate matrix
\begin{equation}  \label{LC} 
\overline{[\Lambda]} = [\Lambda]_{compressed} =    \left(
\begin{array}{ll }   0_{2\times 2}  & I_{2\times 2}   \\   -I_{2 \times 
2}   & R_{2 \times 2} \end{array} \right)
\end{equation}
which characterizes the almost-Poisson structure in 
$\overline{P} = P_{compressed} =  \{ (x,y, u_1, u_2) \}$ with 
\begin{equation}
\overline{H} = H_{compressed} = \frac{1}{2} \,\left( u_1^2 + \frac{u_2^2}{1 + 
x^2}  \right) 
\end{equation}
as the reduced Hamiltonian. It follows that the reduced equations of motion are
$$
\left[ \begin{array}{l} \dot{x}\\ \dot{y} \\ \dot{u}_1 \\ \dot{u}_2    
\end{array} \right]  \, =
\,  \left(
\begin{array}{ll }   0_{2\times 2}  & I_{2 \times 2}   \\   -I_{2 
\times 2}   & R_{2 x 2} \end{array} \right) \, \left[ \begin{array}{l} 
\overline{H}_{x}\\ \overline{H}_{y} \\ \overline{H}_{{u}_1} \\ \overline{H}_{u_2}    
\end{array} \right]
$$
or
$$
\dot{x} = u_1 \,, \,  \dot{y} = \frac{u_2}{1 + x^2} \,,\,
\dot{u}_1 = - \overline{H}_x + R_{12} \overline{H}_{u_2} \,,  \, \dot{u}_2 = 
- \overline{H}_y  - R_{12} \overline{H}_{u_1}
$$
A ``miraculous'' cancelation takes place in the $\dot{u_1}$ equation, 
and the system
is actually
\begin{equation} \label{reducedsystem}
\dot{x} = u_1 \,, \,  \dot{y} = \frac{u_2}{1 + x^2} \,,\,
\dot{u}_1 = 0 \,,\, \dot{u}_2 =  \frac{x}{1 + x^2} u_1 u_2 
\end{equation}
This cancelation has a reason.  The Lagrangian is invariant under
the 1-parameter group $ (x,y,z) \rightarrow (x + \epsilon, y, z)$ and 
the
generator  $\partial/\partial x$ is an admissible vectorfield.  By the 
non-holonomic
Noether theorem \cite{Arnold}, the momentum  $p_x = u_1 = \dot{x}$ is 
conserved.
The 2 degrees of freedom compressed system separates and can be 
integrated
by quadratures:
$$
\ln(u_2) = \int \, a \frac{x_o + at}{1 + (x_o + at)^2} \, dt = 
\frac{1}{2} \ln ( 1 + 
(x_o + at)^2) + {\rm const.} \ \ \ {\rm Thus,} 
$$
$$  u_2 = A \sqrt{ 1 + (x_o + at)^2 }\ \ \,, \ \ \ 
   y = y_o + \int_0^t A \frac{\sqrt{ 1 + (x_o + at)^2 }}{1 + (x_o + 
at)^2} \,dt
$$
and we reconstruct the  z-fiber dynamics via the constraint 
equation:
$$
z = z_o + \int_0^t (x_o + at) \frac{A \sqrt{ 1 + (x_o + at)^2 }}{1 + 
(x_o + at)^2}  dt \ . 
$$

The almost-symplectic form in the compressed space $T^*S$, $S = \R^2 = \{ 
x,y \}$, is the $2$-form 
$\overline{\Omega} = du_1 \wedge dx + du_2 \wedge dy + R_{12} dx \wedge dy$ ,  
$ R_{12} = - \,\frac{x u_2}{ 1 + x^2}$ 
and so 
$d \overline{\Omega} = - \frac{x}{1 + x^2} \, du_2 dx dy$ .  

Let us  investigate if there is a function $f: T^*S \rightarrow \R$ such that $f \overline{\Omega}$ is 
closed.
It looks simpler to try the {\it ansatz} $f = f(x)$ only.
The condition  $df \wedge \overline{\Omega} + f  d \overline{\Omega}$ = 0 leads to
\begin{equation} \label{f}
[f'(x) + f \frac{x}{1 + x^2}] dx du_2 dy  = 0 \ , \ {\rm so \ that} \ \ 
f = \frac{A}{\sqrt{1 + x^2}} \,\,.
\end{equation}
The significance of this observation is that the compressed system 
(\ref{reducedsystem})
is Hamiltonian in the time scale $s$ such that $ dt/ds = \sqrt{1 + x^2}$ ,  
with the same Hamiltonian $\overline{H} = H_{compressed}$ and bona-fide symplectic form  $f \overline{\Omega}$. 
Now, the corresponding Poisson structure in $T^*S$  is $ \frac{1}{f} [\Lambda]_{compressed}$ (refer to (\ref{LC})) 
and one is tempted to guess that the conformally changed bivector in $P$ given by $\frac{1}{f} [\Lambda]_{\rm moving}$ (refer to (\ref{QP}))  
satisfies Jacobi. But as we shall see below, {\it this is   not  the case !}

\section{ Non-Jacobi for the constrained almost-Poisson } 

Let us now take a closer look on the algebraic properties of the 
almost-Poisson 
structure on the constrained space $P$. 
In order to compute the  
Schouten-Nijenhuis bracket of the almost-Poisson bivector $\Lambda$ with 
itself, 
we first rewrite $\Lambda$ given by (\ref{QP}) as a matrix in a 
coordinate basis. Choosing $\{x,y,z,u_1,u_2\}$ as coordinates, 
we write $\Lambda_{\{ x,y,z,u_1,u_2\} }$ as 
$$     \left(
\begin{array}{lll}   0_{2\times 2} & 0_{2 \times 1}
& I_2 \\ 0_{1 \times2} & 0_{1 \times 1 } & L_{1 \times 2} \\   
-I_{2 \times 2} &  -L_{2 \times 1}^T  & R_{2 \times 2} \end{array} 
\right)
$$
where, as before, $R_{2 \times 2}$ is the antisymmetric matrix with 
$R_{12} = -xv/(1+x^2)$  
and now 
$$
L_{1\times 2} = ( \ 0 \ \ x \ ) 
$$
in such a way that we can compute the self S-N bracket $[ \Lambda , 
\Lambda ]$ using:  
\begin{equation} \label{S-N} 
[ \Lambda , \Lambda ]^{IJK} = \Lambda^{LK}\partial_L\Lambda^{IJ} + 
\Lambda^{LI}\partial_L\Lambda^{JK} + \Lambda^{LJ}\partial_L\Lambda^{KI} 
\end{equation} 
where the summation convention is subtended. We then get 
\begin{equation} \label{R1}
[ \Lambda , \Lambda ] = \left(\frac{2}{1+x^2}\right)\left( x\partial_y 
- \partial_z \right) \wedge\partial_{u_1}\wedge\partial_{u_2}  \ \ .
\end{equation} 
One can check explicitly that there is no vector field $E$ satisfying the first of the equations for 
the existence of an associated Jacobi structure \cite{Marle} (the second one is given by (\ref{Jac2})) :  
\begin{equation} \label{Jac}
[ \Lambda , \Lambda ] =  2E\wedge\Lambda \ . 
\end{equation} 
Thus no Jacobi structure, in the sense of Marle, exists for this bi-vector (or any 
conformal one as well). Notice that we can rewrite 
$[ \Lambda , \Lambda ] = \left(\frac{2}{1+x^2}\right) 
e^{\perp}\wedge\partial_{u_1}\wedge\partial_{u_2}$ ,   
where $e^{\perp} = x\partial_y - \partial_z$ is a vector orthogonal 
to $e_1$ and $e_2$ and hence to the distribution.
Remark that  $\Lambda / f$ , where $f$ is given by 
(\ref{f}), is really ``$z$-almost'' Poisson. A calculation
in the same lines shows that the only nonvanishing entries of 
$[\Lambda / f , \Lambda / f]$ are the permutations of  $(3,4,5)$, 
namely
$ -2(1+x^2) \partial_{z} \wedge \partial_{u_1} \wedge \partial_{u_2}$ .  
(``almost'' is really almost)!

In order to better appreciate the Non-Jacobi result for the almost-Poisson
structure, 
let us now choose a different moving frame and co-frame adapted to the 
contact distribution. 
Specifically, we choose an orthonormal set with respect to the euclidean metric: 
\begin{equation}
e_1 = \partial_x  \ , \ e_2 = \frac{\partial_y + 
x\partial_z}{\sqrt{1+x^2}} \ , \ 
e_3 = \frac{\partial_z - x\partial_y}{\sqrt{1+x^2}} 
\end{equation}
whose dual set is ``itself'': 
\begin{equation}
\epsilon_1 = dx  \ , \ \epsilon_2 = \frac{dy + xdz}{\sqrt{1+x^2}} \ , \  
\epsilon_3 = \frac{dz - xdy}{\sqrt{1+x^2}} = 
\frac{\omega}{\sqrt{1+x^2}} \,\,\,.
\end{equation}
This frame does not respect the natural $z$-fibration, but it's still 
true that  
$
\omega (e_1) = \omega (e_2) = 0 
$. 
However, the Lie algebra is now modified :
$ 
[ e_1 , e_2 ] = \left(\frac{1}{1+x^2}\right)e_3 \ , \ [ e_2 , e_3 ] = 0 
\ , \ [ e_3 , e_1 ] = \left(\frac{1}{1+x^2}\right)e_2 
$ , 
but the hamiltonian $H$ corresponding to the euclidean 
metric is again euclidean: 
$
H = (u_1^2 + u_2^2 + u_3^2)/2 
$ ,  
so that the condition $\partial H / \partial u_3 = 0$ yields the 
simpler equation
\begin{equation} \label{zero}
u_3 = 0 
\end{equation}
for the definition of $P$ (generally, we have $u_{\alpha} = 0$ 
for orthonormal frames). 

From the theory of moving frames \cite{KR},  
the Poisson bi-vector on $T^*Q$ can be written as 
\begin{equation}
\hat{\Lambda} = e_I^*\wedge\partial_{u_I} + 
R_{IJ}\partial_{u_I}\wedge\partial_{u_J}
\end{equation} 
where $R_{IJ}$ is given by (\ref{SM}), 
$\partial_{u_I}$ is the ``vertical dual'' to $du_I$ and 
$e_I^*$ 
is the  correct  lift of the base vector $e_I$ to $T(TQ)$ \cite{KR}. 
Alternatively, we can 
rewrite this full bi-vector using 
Darboux quasi-coordinates as 
$\hat{\Lambda} = e_I\wedge\partial_{u_I}$ .
For an orthonormal adapted moving frame, from (5) we get that the almost-Poisson bivector on 
the constrained space $P$ is: 
\begin{equation} \label{LamP1}
\Lambda = {e}_i\wedge\partial_{u_i} + 
\tilde{R}_{ij}\partial_{u_i}\wedge\partial_{u_j} 
\end{equation}
where $\,\,\, \tilde{} \,\,\, $ means evaluating at $P$, in this case: 
\begin{equation} \label{RP}
\tilde{R}_{IJ}  = - u_i\epsilon_i [e_I,e_J] \ .  
\end{equation}

For the contact system with orthonormal moving frame, 
we have simply  
\begin{equation} \label{Lambda}
\Lambda = {e}_1 \wedge\partial_{u_1} + 
{e}_2 \wedge\partial_{u_2}
\end{equation}
where $\partial_{u_i}$ is the ``vertical dual'' to $\epsilon_i$ : 
$
\partial_{u_1} = \partial_{p_x}  \ , \ \partial_{u_2} = 
\frac{\partial_{p_y} + x\partial_{p_z}}{\sqrt{1+x^2}} \ , \ 
\partial_{u_3} = \frac{\partial_{p_z} - x\partial_{p_y}}{\sqrt{1+x^2}}  
$ . 
The self S-N bracket of $\Lambda$ can be easily computed using the 
decomposition formulas 
for the S-N bracket of bivectors. We get : 
$$
[ {e}_1 \wedge\partial_{u_1} + {e}_2 \wedge\partial_{u_2} , 
{e}_1 \wedge\partial_{u_1} + {e}_2 \wedge\partial_{u_2} ]    = 
[ {e}_1 \wedge\partial_{u_1} , {e}_1 \wedge\partial_{u_1} ] + 
[ {e}_2 \wedge\partial_{u_2} ,  {e}_2 \wedge\partial_{u_2} ] + 
2[ {e}_1 \wedge\partial_{u_1} , {e}_2 \wedge\partial_{u_2} ]  
$$
It's easy to see that the first two terms on the right vanish. The 
third one decomposes as  
$$ 
[ {e}_1 \wedge\partial_{u_1} , {e}_2 \wedge\partial_{u_2} ] = 
[ {e}_1  , {e}_2 ] \wedge\partial_{u_1}\wedge\partial_{u_2} - 
{e}_1 \wedge [ {e}_2 , \partial_{u_1} ] \wedge\partial_{u_2} -  
{e}_2 \wedge [ {e}_1 , \partial_{u_2} ] 
\wedge\partial_{u_1} + {e}_1 \wedge {e}_2  \wedge [ \partial_{u_1} , \partial_{u_2} ] 
$$ 
Since the only nontrivial Lie bracket is 
$
 [ e_1 , e_2 ] = 
\left(\frac{1}{1+x^2}\right)e_3 
$ 
we are left with 
\begin{equation} \label{R2}
[ \Lambda , \Lambda ] =  
\left(\frac{2}{1+x^2}\right)e_3\wedge\partial_{u_1}\wedge\partial_{u_2} \ ,  
\end{equation}
which is equivalent to the  result previously obtained 
(\ref{R1}). It's now even simpler 
to verify that no Jacobi structure \cite{Marle}\cite{Marle2} is possible for $\Lambda$, just 
compare  (\ref{Jac}), (\ref{Lambda}) and (\ref{R2}).

Let us consider a still simpler, even more symmetrical case. 
Namely, on the contact system, we consider a metric which is invariant 
under the full Heisenberg group. 
Again, taking (\ref{frame}) as moving frame  we have the Heisenberg 
algebra (\ref{Heisen}). 
The simplest Heisenberg-invariant metric is thus given by the kinetic 
energy
$ T = \frac{1}{2}(v_1^2 + v_2^2 + v_3^2) $ , 
where $v_i$ is given by the identification 
$ v_1e_1 + v_2e_2 + v_3e_3 = \dot{x}\partial_x + \dot{y}\partial_y + 
\dot{z}\partial_z $ , 
which gives: 
\begin{equation} \label{HT}  
T =  \frac{1}{2}(\dot{x}^2 + (1 + x^2)\dot{y}^2 + \dot{z}^2 - 
2x\dot{y}\dot{z}) \ . 
\end{equation}  
In other words, for such metric $2T$, the Heisenberg moving frame 
(\ref{frame}) is orthonormal. 
In terms of the hamiltonian, we have that the co-frame (\ref{coframe}) 
is also orthonormal, and thus 
\begin{equation} \label{simple}   
H = \frac{1}{2}(u_1^2 + u_2^2 + u_3^2) = \frac{1}{2}( p_x^2 + p_y^2 + 
(1 + x^2)p_z^2 + 2xp_yp_z ) 
\end{equation} 
where the relation between the u's and the p's is the one given earlier 
in (\ref{change}), section 3.
Once again,  the almost-Poisson bivector is given by 
\begin{equation}  
\Lambda = e_1\wedge\partial_{u_1} + e_2\wedge\partial_{u_2} \ , \ \ {\rm and \ \ thus} \ \ \  
 [\Lambda , \Lambda ] = 2e_3\wedge\partial_{u_1}\wedge\partial_{u_2} \ . 
\end{equation} 
Therefore, no Jacobi structure \cite{Marle}\cite{Marle2} exists on $P$ for 
this Schaft-Maschke \cite{SM} bivector.  

\section{The compressed almost-Poisson structure \\ is not conformally symplectic in general} 

We've  seen earlier in section $5$ that the compressed system for the contact 
nonholonomic system with euclidean 
metric is conformally symplectic.
However  such conformal symplectic structure (or any 
Jacobi structure) 
on the compressed system does not exist for generic metrics. To see this, 
consider 
a $z$-invariant hamiltonian of the general form 
\begin{equation}
H = \gamma_{ij}u_iu_j + V(x,y) \ , \  \gamma_{ij} \equiv 
\gamma_{ij}(x,y) = 
\gamma_{ji}(x,y)  \  .
\end{equation}
The condition $\partial H/\partial u_3 = 0$ yields
$u_3 = - (\gamma_{13}u_1 + \gamma_{23}u_2)/\gamma_{33} = - R_{12}$ ,   
so that the compressed almost-Poisson bivector can be written as 
\begin{equation}  
\overline{\Lambda} = \partial_x\wedge\partial_{u_1} + 
\partial_y\wedge\partial_{u_2} + 
\left(\frac{\gamma_{13}u_1 + \gamma_{23}u_2}{\gamma_{33}}\right)
\partial_{u_1}\wedge\partial_{u_2}  \,\,.
\end{equation} 
Since $\{x,y,u_1,u_2\}$ is a coordinate system for $T^*S$, we can apply 
formula (\ref{S-N}) directly to obtain 
$[ \ \overline{\Lambda} , \overline{\Lambda} \ ] = - 
2\left(\frac{\gamma_{13}}{\gamma_{33}}\right)\partial_x\wedge\partial_{u_1}\wedge\partial_{u_2} -
2\left(\frac{\gamma_{23}}{\gamma_{33}}\right)\partial_y\wedge\partial_{u_1}\wedge\partial_{u_2}$  . 
On the other hand, a general vector field on $T^*S$ has the form : 
\begin{equation} \label{E1}
E = \alpha\partial_x + \beta\partial_y + \mu\partial_{u_1} + 
\nu\partial_{u_2} 
\end{equation} 
so that the condition 
$ 
[ \ \overline{\Lambda} , \overline{\Lambda} \ ] = 2E\wedge\overline{\Lambda}
$
is satisfied for 
\begin{equation} \label{E2}
\nu = - \gamma_{13}/\gamma_{33} \ , \ \mu = \gamma_{23}/\gamma_{33} \ .
\end{equation}
For $E$ to be the vector field of a Jacobi structure \cite{Marle}\cite{Marle2}, it is 
also necessary that 
\begin{equation} \label{Jac2}  
[ E , \overline{\Lambda} ] \equiv {\mathcal{L}}_E(\overline{\Lambda} ) = 0  \ .
\end{equation} 
A simple computation gives  
$[ E , \overline{\Lambda} ] = 
- \left(\frac{\partial}{\partial 
x}\left(\frac{\gamma_{13}}{\gamma_{33}}\right) + 
\frac{\partial}{\partial 
y}\left(\frac{\gamma_{23}}{\gamma_{33}}\right)\right)
\partial_{u_1}\wedge\partial_{u_2}$ and thus   
 \begin{equation} \label{restriction} 
\frac{\partial}{\partial x}\left(\frac{\gamma_{13}}{\gamma_{33}}\right) 
+ 
\frac{\partial}{\partial y}\left(\frac{\gamma_{23}}{\gamma_{33}}\right) 
= 0
\end{equation}
is a necessary condition for the existence of a Jacobi structure on the 
compressed system in $T^*S$. 
Of course, the hamiltonians (\ref{hamiltonian}) and (\ref{simple}), 
obtained from the euclidean and the Heisenberg metrics, (\ref{everybody}) and (\ref{HT}) respectively, 
satisfy the above condition. More generally, in order to verify this 
condition, one can substitute for the original
metric elements $g_{ij}(x_1,x_2)$ in a basis $x_1 = x , x_2 = y , x_3 = 
z$, with the relations 
\begin{equation}
\frac{\gamma_{13}}{\gamma_{33}} = 
\frac{g_{12}g_{23} - g_{13}g_{22} - x_1(g_{12}g_{33} - g_{13}g_{23})}
{g_{11}g_{22} - g_{12}^2 -2x_1(g_{11}g_{23} - g_{12}g_{13}) + 
x_1^2(g_{11}g_{33} - g_{13}^2)} \ , 
\end{equation}
\begin{equation}
\frac{\gamma_{23}}{\gamma_{33}} = 
\frac{g_{11}g_{23} - g_{12}g_{13} - x_1(g_{11}g_{33} - g_{13}^2)}
{g_{11}g_{22} - g_{12}^2 -2x_1(g_{11}g_{23} - g_{12}g_{13}) + 
x_1^2(g_{11}g_{33} - g_{13}^2)} \ . 
\end{equation} 

Clearly, (\ref{E1}), (\ref{E2}) and (\ref{restriction}) mean that any possible 
Jacobi structure on this compressed system 
comes from a conformal symplectic structure.

\section{Conclusions} 

We've seen that the failure of the Jacobi identity for the 
Schaft-Maschke \cite{SM} almost-Poisson structure on the constrained  submanifold
$P \subset T^*Q$ of a nonholonomic system 
cannot be cirvumvented by the introduction of an  
associated Jacobi structure \cite{Marle}\cite{Marle2}, even in the simplest cases. 
It remains to be confirmed whether 
this ``no-go'' result is generic.

These examples also suggest that pehaps the almost-Poisson structure can be most useful when 
there is a principal
bundle structure $G \rightarrow Q \rightarrow S$, and all data are 
equivariant with respect to
the group $G$. In the contact case, $z$-invariant Lagrangians. 
In this case, the degenerate almost-Poisson bracket in $P$ projects over a 
non-degenerate almost-Poisson bracket in $T^*S$, which in favorable cases (not all) is 
conformally symplectic, or Jacobi. It remains  to be found a geometrical interpretation for these  cases. 

More generally, it also remains to be determined whether or when such a conformally symplectic reduction can be applied for less simple nonholonomic systems. Work in this direction is under way and shall be reported elsewhere.


\end{document}